\documentclass[11pt,a4paper]{article}
\usepackage{amsmath, amsthm, amssymb}
\usepackage{enumerate}
\usepackage{multicol}
\usepackage{epsfig} 
\usepackage{graphicx}
\usepackage[top= 3.0cm, bottom=3.0cm, left=2.5cm, right=2.5cm]{geometry} 
\usepackage[latin1]{inputenc}
\usepackage[T1]{fontenc} 
\usepackage[english]{babel}
\usepackage{dsfont}
\usepackage{cite} 
\usepackage{setspace}
\usepackage{mathtools}
\usepackage{amsthm}
\usepackage{bm}

\numberwithin{equation}{section}

\title{\huge Investigating total collisions
of the \\ Newtonian N-body problem on shape space}

  \author{
         Paula Reichert\thanks{Mathematisches Institut, Ludwig-Maximilians-Universit\"at M\"unchen. E-mail: reichert@math.lmu.de.}\\
          \mbox{}\\
          }

\date{December 10, 2020\vspace{0.7cm}\\
Forthcoming in: \textit{Foundations of Physics}.
}

\begin{document}
\maketitle
\thispagestyle{empty}

\begin{abstract}

\noindent We analyze the points of total collision of the Newtonian gravitational system on shape  space (the relational configuration space of the system). While the Newtonian equations of motion, formulated with respect to absolute space and time, are singular at the point of total collision due to the singularity of the Newton potential at that point, this need not be the case on shape space where absolute scale doesn't exist. We investigate whether, adopting a relational description of the system, the shape degrees of freedom, which are merely angles and their conjugate momenta, can be evolved through the points of total collision. Unfortunately, this is not the case. Even without scale, the equations of motion are singular at the points of total collision (and only there). This follows from the special behavior of the shape momenta. While this behavior induces the singularity, it at the same time provides a purely shape-dynamical description of total collisions. By help of this, we are able to discern total-collision solutions from non-collision solutions on shape space, that is, without reference to (external) scale. We can further use the shape-dynamical description to show that total-collision solutions form a set of measure zero among all solutions. \end{abstract}

\newpage

\section{Introduction}
\pagenumbering{arabic}

Is the Big Bang the beginning of Everything? Or did the universe evolve through the Big Bang which would then provide a single instant in the eternal evolution of the world? In what follows, we want to analyze the points of total collision of the Newtonian $N$-body system -- the Big Bang of the Newtonian universe -- on shape space, the relational configuration space of the system.

It is well-known that, in the usual description with respect to absolute space and time, the Newtonian solutions end at or begin at the point of total collision. This is due to the singularity of the Newton potential at that point. Being inversely proportional to the inter-particle distances, the Newton potential diverges at the point at which the inter-particle distances go to zero. As a consequence, also the Newtonian vector field and the Newtonian equations of motion are singular at that point. This is why the Newtonian solutions, as described with respect to absolute space, cannot be continued through the point of total collision. 

In what follows we want to investigate whether the Newtonian solutions can be evolved through the points of total collision on shape space. On shape space, absolute scale is no longer part of the description, only angles (i.e. shape degrees of freedom) remain. Do these shape degrees of freedom evolve (uniquely) through the points of collision? Unfortunately, this is not the case. We will show why, although there exists a unique description of the Newtonian system on shape space, a description, that is, which is free of scale, there is still a singularity at the points of total collision. This has to do with the special behavior of the shape momenta at that point. While this means that, even on shape space, solutions cannot be continued through the points of total collision, we, at least, obtain a purely shape-dynamical description of total collisions, which allows us to specify the points of collision without reference to scale.

Throughout the paper, we consider the Newtonian model of zero total energy $E=0$, zero linear momentum $\textbf{P}=0$ and zero angular momentum $\textbf{L}=0$ (where $E$, $\textbf{P}$ and $\textbf{L}$ can be fixed because they are conserved quantities of motion), the so-called $E=\textbf{P}= \textbf{L}=0$ Newtonian universe. In order to be able to compute everything explicitly, we later restrict the discussion to the three-body model. A similar, though more abstract discussion, however, can be made for the N-body problem. In particular, it can be shown that, also for $N$ particles, the shape Hamiltonian is a sum of squares of the shape momenta, the inverse of which enters the shape velocities, which, at the points  of total collision on shape phase space, diverge. Hence, also for the relational $N$-body problem, there will be a singularity at the points of total collision.

The discussion is essentially based on a reformulation of the Newtonian dynamics on shape space, the relational configuration space of the system, where absolute position, orientation and size are no longer part of the description. Instead, the system's evolution is fully determined by the evolution of shapes: triangle shapes in the case of three particles, where each triangle shape is specified by two angles. For the three-particle Newtonian gravitational system with $E=\textbf{P}= \textbf{L}=0$, a relational formulation with scale, but without absolute position and orientation has been obtained by Montgomery [2002]. A fully relational (Hamiltonian) formulation of this system on scale-invariant shape phase space has eventually been obtained by Barbour, Koslowski, and Mercati ([2013], [2015]). 

It is interesting that a shape-dynamical formulation of the Newtonian dynamics exists although the system is not scale-invariant in the first place. We show how this works and how all remnants of scale are hidden in the time-dependence of the Hamiltonian vector field on shape space. While the shape vector field is still singular at the points of total collision, we, at least, detect a purely shape-dynamical description of total collisions. This allows us to discern total-collision solutions from non-collision solutions without reference to scale. We can also use this shape-dynamical description of total collisions to show that the set of total-collision solutions forms a set of measure zero among all solutions.

In what follows, we first analyze the $E=\textbf{P}= \textbf{L}=0$ Newtonian gravitational system in its formulation with respect to absolute space and time. We will study both the long-time behavior and the behavior at the points of total collision. This will constitute section 1. In section 2, we derive the Newtonian dynamics on shape phase space, the relational phase space of the system. In section 3, we study the points of total collisions on shape space. We show that, due to the special behavior of the shape momenta, the shape degrees of freedom cannot be evolved through the points of total collision. Still, from the behavior of the shape momenta we obtain a purely shape-dynamical description of total collisions. We conclude with a discussion of the result.

\section{Dynamics in absolute space and time}
Throughout the paper, the term `Newtonian universe' refers to the model of $N$ particles moving through infinite, three-dimensional Euclidean space and attracting each other according to the Newtonian gravitational force law. In what follows, we consider the $E=\textbf{P}= \textbf{L}=0$ Newtonian universe, i.e., the Newtonian model of zero total energy, $E=0$, zero total linear momentum, $\textbf{P}=0$, and zero total angular momentum, $\textbf{L}=0$.

\subsection{The long-time behavior of the system}

It has been an early result of the analysis of the gravitational $N$-body system that the long-time behavior of the Newtonian universe is governed by the {Lagrange-Jacobi equation}, 
\begin{equation} \ddot{I}= 4E -2V_{N}. \end{equation}
This equation connects the second time derivative of the moment of inertia $I$ to the total energy $E$ and the potential energy $V_{N}$ of the system. The moment of inertia in the center-of-mass frame (which we can use without loss of generality because the system is invariant under total translations) is of a simple form, \begin{equation} I=\sum_{i=1}^N m_i \textbf{q}_i^2.\end{equation} 
Note that, by definition, the moment of inertia $I$ is an intrinsic measure of the size of the system.

The Newton potential is defined to be
\begin{equation} V_{N} =  -\frac{1}{2} \sum^N_{i\neq j; i,j=1} \frac{Gm_im_j}{|\textbf{q}_i-\textbf{q}_j|} \end{equation} 
and $E=T+V_{N}$ is the total energy, with $T=\sum_{i=1}^N \textbf{p}_i^2/2m_i$ being the kinetic energy of the system.

Since we consider the model of zero total energy, $E=0$, and since the potential energy is strictly negative, $V_{N}<0$, the Lagrange-Jacobi equation tells us that the second time derivative of $I$ is strictly positive:
\begin{equation} \ddot{I}>0.\end{equation}
This condition implies that the $I$-curve, if the solution exists for all times, is concave upwards with a minimum $I_{min}$ at some moment $t_0$ and with $I$ increasing towards infinity in both time directions away from that point, $I\to \infty$ as $t\to \pm \infty$. In the case of a total collision, where the Newtonian solution ends at some point (the point of total collision) due to the singularity of the vector field at that point, the Lagrange-Jacobi equation implies that $I_{min}=0$ at $t_0$ (expressing the fact that the particles collide at some moment) while the $I$-curve is concave and goes to infinity as $t\to \infty$ or $t\to -\infty$ (which is the same since the system is time-symmetric). Let from now on, without loss of generality, $t_0=0$.

Since the moment of inertia $I$ is an intrinsic measure of the size of the system, we learn from the Lagrange-Jacobi equation that, in an eternal evolution, at one moment, the particles are closest while they become spatially more and more separated in both time directions away from that point. If a Big Bang is part of such an eternal evolution -- assume, for one moment, that the shape degrees of freedom can be evolved through the singularity of a total collision --, it has to occur at the mid-point of the evolution, when the spacing between the particles is minimal, $I=I_{min}$, and where, in the case of a total collision (a Big Bang of the Newtonian universe), $I_{min}=0$. Only at that point, when $I=I_{min}$, the Big Bang can be part of an eternal evolution (if at all) of the $E=0$ Newtonian universe.

From the Lagrange-Jacobi equation we further learn that, for the $E=0$ Newtonian universe, there exists a quantity which is strictly monotonic along the trajectories: the {`dilational momentum'} \begin{equation} D=\sum_{i=1}^N \textbf{q}_i\cdot\textbf{p}_i.\end{equation} Expressed with respect to $I=\sum_i m_i\textbf{q}_i^2$, we find that it is essentially its first time derivative,
\begin{equation} D=1/2 \dot{I}. \end{equation} Since $\ddot{I}>0$, it follows that $\dot{I}$ is monotonically increasing along all trajectories. Hence, also $D=1/2 \dot{I}$ is monotonically increasing. Thus we can use $D$ in order to specify the minimum of the $I$-curve. Since $D\sim \dot{I}$, it follows that $I=I_{min}$ if and only if
\begin{equation} D=0. \end{equation}
Given that $D$ is monotonic, we can use it to parametrize the trajectories. Once we do that, we call it `{internal time}' and write $\tau=D$. 
In fact, it will be this internal time parametrization which allows us to reformulate the Newtonian dynamics, which is not invariant under scalings/dilations, on scale-invariant shape space (cf. Barbour et al. [2013]).

Let me add one remark. The time-asymmetric processes of overall contraction and expansion of the system defines two {gravitational arrows of time}. Following Barbour et al. ([2013], [2015]) we call `{past}' the mid-point of the evolution when the particles are closest, $\tau=0$, while we say there are two `{futures}' in both time directions away from it, as $\tau\to \pm \infty$. If there is a total collision at the mid-point of the evolution (which Barbour et al. also call Janus point), we say there is a Big Bang in the common past of the Newtonian universe.

\subsection{Total collisions} In order to study whether the shape degrees of freedom can be evolved through the Newtonian singularity, we need to analyze the Newtonian dynamics at the points of total collision. 

In what follows, I introduce the standard definition of a total collision. We say that a total collision occurs if and only if the moment of inertia vanishes: 
\begin{equation}I=0.\end{equation}

Total collisions have been discussed in the context of the three-body problem early in the mathematical literature. The existence of solutions ending at, respectively starting at a total collision has been shown by Lagrange and Euler.\footnote{Cf. Moeckel ([1981], [2007]) for a historical introduction.} Lagrange showed that, if there are three particles of equal masses forming an equilateral triangle and they are released with zero initial velocity, they will collide. We will call this particular spatial configuration plus its reflected version (the reflected equilateral triangle) the two {`Lagrange configurations'}. Euler, in turn, showed that, if there are three particles of equal masses aligned with one particle centered between the other two and they are released with zero initial velocity, they will collide. The respective configurations are called the {`Euler configurations'}. There are three Euler configurations, one for each possibility to center one particle in-between the other two.

Today, some general properties of total collisions are known. Sundman [1909] has shown that, in order for a total collision to occur, the total angular momentum needs to be zero, $\textbf{L}=0$ (if $\textbf{L}\neq 0$, $I$ is bounded away from zero by some positive constant: $I\ge I_0$ with $I_0>0$). Later Saari [1984] has shown that, as the particles approach a total collision at time  $t_0=0$, they form a central configuration and their position vectors $\textbf{q}_i(t)$ behave as $t^{2/3}$. Consistently, Moeckel ([1981], [2007]) identifies the central configurations as the `rest points' of the collision manifold. This specifies the points of total collision on shape space. For three particles, there exist five points of central configuration on shape space: the three Euler and two Lagrange points. These are the only points on shape space at which a total collision may occur. 

To gain some intuition about total collisions due the attractive gravitational force, let us sketch the asymptotic behavior of the particles as they approach a total collision.\footnote{The following paragraphs follow Saari's discussion of the matter in his paper on central configurations.} Saari [1984] shows that, as the particles approach a total collision at $t_0=0$, their position vectors behave as $\textbf{q}_i(t)=\textbf{a}_i t^\alpha$ for some $\alpha$, where $\textbf{a}_i$ is a vector constant. To be precise, he shows that there exist positive vector constants $\textbf{A}_i, \textbf{B}_i$ such that, for sufficiently small $t$ (in particular, for $t\to 0$), 
\begin{equation}  \textbf{A}_i t^\alpha \le \textbf{q}_i(t) \le \textbf{B}_i t^\alpha.\end{equation}
From this result it follows that, in the limit $t\to 0$, the colliding particles form a central configuration (see below). Moreover, we can even specify the $\alpha$: in the limit $t\to 0$, it is $\alpha= 2/3$. Hence, the particles' position vectors behave as \begin{equation}\textbf{q}_i(t)=\textbf{a}_i t^{2/3}.\end{equation} 
To see this, reconsider the Newtonian gravitational potential (2.3):
\[ V_{N} =  -\frac{1}{2} \sum_{i\neq j} \frac{Gm_im_j}{|\textbf{q}_i-\textbf{q}_j|}. \]
This potential forms part of the Newtonian law of gravitation which determines the acceleration $\ddot{\textbf{q}}_i$ of the $i$'th particle (with mass $m_i$) as follows:
\begin{equation} m_i\ddot{\textbf{q}}_i =\frac{\partial V_{N}}{\partial \textbf{q}_i}= - \sum_{i\neq j} \frac{Gm_im_j (\textbf{q}_i-\textbf{q}_j)}{|\textbf{q}_i-\textbf{q}_j|^3}. \end{equation}

This is a complicated equation, but sometimes it attains a simpler form. This is the case for central configurations. A configuration is called a {`central configuration'} if, at some moment, the center-of-mass position vector $\textbf{q}_i$ of each particle is in line with its acceleration vector $\ddot{\textbf{q}}_i$. That is, if and only if $ \forall i=1,...,N$:
\begin{equation} \lambda \textbf{q}_i =  \ddot{\textbf{q}}_i,\end{equation}
respectively, with (2.10):
\begin{equation} \lambda\textbf{q}_i = \frac{1}{m_i} \frac{\partial V_{N}}{\partial \textbf{q}_i}.\end{equation}
Here $\lambda= \lambda(t)$ is some common scalar factor of proportionality. Hence, if the particles form a central configuration, the system mimics a central force problem. 

Using Saari's result, namely that, in the limit $t\to 0$, the position vectors behave as $\textbf{q}_i(t)=\textbf{a}_i t^\alpha$ for some $\alpha$, the Newtonian force law (2.10) turns into
\begin{equation}  \textbf{a}_i \alpha(\alpha-1)t^{\alpha-2}= - \sum_{i\neq j} \frac{Gm_j (\textbf{a}_i-\textbf{a}_j)}{|\textbf{a}_i-\textbf{a}_j|^3}t^{-2\alpha}. \end{equation}
This can be fulfilled if and only if $\alpha=2/3$. Consequently, $\textbf{q}_i(t)=\textbf{a}_i t^{2/3}$ and the system forms a central configuration with $\lambda(t)=  \alpha(\alpha-1)t^{-2} = -2/9 t^{-2}$, where the functional form of $\lambda$ follows from differentiating $\textbf{q}_i(t)=\textbf{a}_i t^\alpha$ twice with respect to $t$:
\begin{equation} \ddot{\textbf{q}}_i= \textbf{a}_i \alpha(\alpha-1)t^{\alpha-2}=\lambda(t)  \textbf{a}_i t^{\alpha}=\lambda(t) \textbf{q}_i. \end{equation}  
That is, in the limit $t\to 0$ (with a total collision at $t_0=0$) the particles form a central configuration and $\forall i=1, ..., N$ there exist vector constants $\textbf{a}_i$ such that $\textbf{q}_i(t) =\textbf{a}_i t^{2/3}$.\\

\subsection{From absolute space to shape space}

From the long-time behavior of the $E=0$ Newtonian universe we learned that, within an eternal evolution, a total collision can occur only at the mid-point $t_0=0$ of the evolution, when the spacing between the particles is minimal: $I=I_{min}$. Of course, $I_{min}=0$ in the special case of a total collision. We also learned that the mid-point $t_0$ of the evolution is determined by $D=0$ (since $D\propto \dot{I}$ and $\dot{I}=0$ determines the minimum of the $I$-curve).

From the behavior near total collisions we learned that a total collision can occur only if the particles form a central configuration. For a system of three particles, there exist five central configurations: two Lagrange and three Euler configurations. Within a relational description, these configurations are represented by five points on shape space: the two Lagrange and three Euler points. 

In what follows, starting from the standard Hamiltonian description on absolute phase space, we will derive the formulation of the dynamics of the three-particle $E=\textbf{P}= \textbf{L}=0$ Newtonian universe on the relational or shape phase space. We will show that, while on absolute phase space the vector field diverges at the point of zero spatial extension ($I_{min}=0$) due to the singularity of the Newton potential, the shape vector field is non-singular at the Euler and Lagrange points at $D=0$ ($I=I_{min}$) for all but a measure-zero set of solutions. Unfortunately, this measure-zero set of solutions is precisely the set of solutions for which a total collision occurs. 

While total collisions cannot be passed, not even on shape space, we obtain from the relational analysis a purely shape-dynamical description of total collisions. That is, we will be able to specify the points of total collision not by reference to scale ($I=0$), but merely in terms of shape degrees of freedom.

\section{Dynamics on shape space}

If we consider a system with symmetries, like the $E=\textbf{P}= \textbf{L}=0$ Newtonian universe which is symmetric with respect to total translations and rotations, the dynamics doesn't have to be formulated on ordinary phase space $\Gamma$. Instead, there exists a unique description of the system and its dynamics on a lower-dimensional space, the {`reduced phase space'} $\Gamma_{0}$.

Mathematically, $\Gamma_{0}$ and the dynamics on $\Gamma_0$ are constructed from $\Gamma$ and the dynamics on $\Gamma$ by a method called symplectic reduction.\footnote{See, for instance, Arnol'd [1989]. For the mathematical details, see also Marsden and Weinstein [1974] or Iwai [1987]. See Montgomery [2002] and Barbour, Koslowski, and Mercati ([2013], [2015]) for the reduction of the Newtonian gravitational system.} For this paper, it suffices to know that the reduced phase space and the reduced Hamiltonian equations of motion are obtained by fixing both the conserved quantities of motion (here $\textbf{P}$ and $\textbf{L}$)  and the connected gauge degrees of freedom (here absolute position and orientation) related to the symmetries of the system. 

\subsection{Reduced phase space $\Gamma_{0}=T^*\mathcal{S}_R$ and shape phase space $T^*\mathcal{S}$}

Within the Newtonian universe, it follows from the conservation of total linear momentum  and total angular momentum, $\{\textbf{P}, H\}=0$ and $\{\textbf{L}, H\}=0$ where $H=T + V_N$ is the Hamiltonian of the system, $\textbf{P}=\sum_i \textbf{p}_i$ total linear momentum and $\textbf{L}=\sum_i\textbf{q}_i\times\textbf{p}_i$ total angular momentum, that the dynamics is invariant under spatial translations and rotations. This follows from Noether's theorem. In the given case, where $\textbf{P}= \textbf{L}=0$, the dynamics is invariant under the full six-dimensional Euclidean group $E(3)=\mathbb{R}^3\times SO(3)$ of spatial translations and rotations, where $\mathbb{R}^3$ denotes the group of translations and $SO(3)$ the group of rotations.\footnote{Note that only if $\textbf{P}=\textbf{L}=0$, the system is invariant under the full Euclidean group. If, e.g., $\textbf{L}\neq 0$, the system is only invariant with respect to one-dimensional rotations (around the axis pointing into the direction of $\textbf{L}$) and, thus, only with respect to a subgroup of $E(3)$. Also only in the given case, if $\textbf{P}=\textbf{L}=0$, the reduced phase space $\Gamma_{0}$ is isomorphic to (and, hence, can be identified with) the cotangent bundle of the reduced configuration space $T^*\mathcal{S}_R$ which we construct below. For details, see the references in the preceding paragraph.} 

One can now construct the reduced phase space $\Gamma_0$ (or at least one representative of it) from phase space $\Gamma\cong\mathbb{R}^{6N}$ by setting $\textbf{P}=\sum \textbf{p}_i=0$ and $\textbf{L}=\sum \textbf{q}_i\times\textbf{p}_i=0$ (thereby fixing the six conserved quantities of motion) and specifying the position and orientation of the system (thereby fixing the gauge degrees of freedom related to translational and rotational symmetry). To do the latter, we fix the center of mass to the origin, $\textbf{Q}_{cm}=\sum_{i=1}^N m_i\textbf{q}_i=0$, and the three off-diagonal components $I_{ij}$ of the (symmetric) center-of-mass inertia tensor to the coordinate axes: $I_{ij}=[\sum_{k=1}^3 m_k (\textbf{q}_k\cdot \textbf{q}_k \mathbb{I} - \textbf{q}_k \otimes \textbf{q}_k)]_{ij}  =0$ (with $i<j; i, j=1,2,3$). Let, to simplify notation, $\textbf{I}_L:= (I_{12}, I_{23}, I_{13})$. That is, we set $\textbf{I}_L=0$. Inserting these conditions into the equations of motion, we obtain the reduced Hamiltonian equations, that is, the equations of motion on $\Gamma_0$. Since we have twelve constraints, $\Gamma_0$ is a space of $6N-12$ dimensions.

An equivalent way to obtain $\Gamma_0$ is by first constructing the reduced configuration space $\mathcal{S}_R$  -- what we call `{shape space with scale}' -- and then determining its cotangent bundle $T^*\mathcal{S}_R$. Just like phase space is the cotangent bundle of configuration space, $\Gamma=T^*Q$, reduced phase space is the cotangent bundle of reduced configuration space, $\Gamma_0= T^*\mathcal{S}_R$.

The reduced configuration space $\mathcal{S}_R$ is obtained from ordinary configuration space $Q\cong\mathbb{R}^{3N}$ by factoring out translations and rotations. To be precise, $\mathcal{S}_R$ is the Riemannian quotient of $Q$ with respect to the Euclidean group $E(3)=\mathbb{R}^3\times SO(3)$, 
\begin{equation} \mathcal{S}_R = \frac{Q}{\mathbb{R}^3\times SO(3)}.\end{equation} 
We call $\mathcal{S}_R$ shape space with scale in order to emphasize that what is left in the description of the system is shapes, that is, angles (or relative distances, respectively) and scale, that is, one quantity measuring the size of the system.

For a system of three particles, $\mathcal{S}_R$ is the `{space of triangles}'. Every point on $\mathcal{S}_R$ determines a distinct triangle, specified by two angles $\psi$ and $\phi$, determining the shape of the triangle, and one scale factor $R$, specifying its size.
Geometrically, $\mathcal{S}_R$ is a cone over the two-sphere $S^2$ (which can be visualized as a collection of two-spheres around the origin with radius $R$). Local coordinates of $\mathcal{S}_R$ are, e.g., spherical coordinates $\psi$, $\phi$, and $R$.

If we further quotient by dilations/scalings, we end up with `{shape space}' $\mathcal{S}$. Shape space $\mathcal{S}$ is the quotient of configuration space $Q$ with respect to the seven-dimensional similarity group $Sim(3) = \mathbb{R}^3\times SO(3)\times\mathbb{R}^+$, that is,
\begin{equation} \mathcal{S}= \frac{Q}{ \mathbb{R}^3\times SO(3)\times\mathbb{R}^+}.\end{equation}
We call $\mathcal{S}$ shape space (or shape sphere, since, geometrically, it is a two-sphere) in order to emphasize that what is left in the description of the system is shapes, that is, angles. 
Note that we have not quotiented by reflections. This is why $S$ is the entire two-sphere $S^2$ and not only the upper half-sphere (the lower half-sphere consists of the reflected triangle shapes). 

For three particles, $\mathcal{S}$ is the {space of triangle shapes}. Every point on $\mathcal{S}$ determines a distinct triangle shape, specified by two angles $\psi$ and $\phi$. Geometrically, $\mathcal{S}$ is represented by a unit two-sphere around the origin. Local coordinates of $\mathcal{S}$ are, e.g., polar coordinates $\psi$ and $\phi$.
Just like $T^*S_R$ is the phase space related to shape space with scale, the cotangent bundle of shape space $T^*\mathcal{S}$ is the `shape phase space'.

\subsection{Local coordinates of $T^*S_R$}

Since the dynamics is invariant with respect to translations and rotations, we can formulate it on the reduced phase space $\Gamma_0=T^*\mathcal{S}_R$. To obtain the reduced Hamiltonian equations of motion, we have to separate the absolute from the relational degrees of freedom. This is achieved by two convenient coordinate transformations. To formulate the dynamics on $\Gamma_0$, we need a canonical set of local coordinates of $T^*\mathcal{S}_R$. In other words, we need a set of translationally and rotationally invariant coordinates plus their canonical conjugates. 

Let, from now on, $N=3$. In that case, $Q\cong\mathbb{R}^9$ and $\Gamma=T^*Q\cong\mathbb{R}^{18}$ and the components of the positions $\textbf{q}_i\in \mathbb{R}^3$ ($i=1,2,3$) and momenta $\textbf{p}_i\in\mathbb{R}^3$ ($i=1,2,3$) form a canonical set of local coordinates of $\Gamma$. In what follows, we first construct the translationally invariant Jacobi coordinates $ \boldsymbol{\rho}_i\in\mathbb{R}^3$ ($i=1,2$) from which, in turn, we obtain the translationally and rotationally invariant Hopf coordinates $\textbf{w}=(w_1, w_2, w_3)$. The three Hopf coordinates are local coordinates of $\mathcal{S}_R$. Together with their canonical conjugates $\textbf{z}=(z_1, z_2, z_3)$ they form a canonical set of local coordinates of $T^*\mathcal{S}_R$.

Starting from the canonical coordinates $\textbf{q}_i\in \mathbb{R}^3$ ($i=1,2,3$) and $\textbf{p}_i\in \mathbb{R}^3$ ($i=1,2,3$) and setting $\textbf{Q}_{cm}=0$ and $\textbf{P}=0$, we obtain the six translationally invariant Jacobi coordinates:
\begin{eqnarray} \boldsymbol{\rho}_1&=&\sqrt{\frac{m_1 m_2}{m_1+m_2}}(\textbf{q}_1-\textbf{q}_2),\nonumber\\
\boldsymbol{\rho}_2&=&\sqrt{\frac{m_3(m_1+ m_2)}{m_1+m_2+m_3}}\bigg(\textbf{q}_3-\frac{m_1\textbf{q}_1 + m_2\textbf{q}_2}{m_1+m_2}\bigg).
\end{eqnarray}
Their conjugate momenta are: 
\begin{eqnarray} \boldsymbol{\kappa}_1 &=& \frac{m_1\textbf{p}_2-m_2\textbf{p}_1}{\sqrt{m_1m_2(m_1+m_2)}},\nonumber\\
\boldsymbol{\kappa}_2 &=&  \sqrt{\frac{(m_1+m_2)}{(m_1+m_2+m_3)m_3}}  \bigg(\textbf{p}_3 -\frac{m_3\textbf{p}_1+m_3\textbf{p}_2}{m_1+m_2}\bigg).
\end{eqnarray}

From these, setting $\textbf{L}=0$ and $\textbf{I}_L=0$, we obtain the three rotationally invariant Hopf coordinates (which have originally been proposed by Hopf in his discovery of the Hopf fibration, see Montgomery [2002]):
\begin{equation} w_1= \frac{|{\boldsymbol{\rho}}_1|^2-|{\boldsymbol{\rho}}_2|^2}{2}, \hspace{1cm}w_2={\boldsymbol{\rho}}_1\cdot{\boldsymbol{\rho}}_2, \hspace{1cm} w_3= {\boldsymbol{\rho}}_1 \times {\boldsymbol{\rho}}_2.\end{equation} 
Their canonical conjugates are:
 \begin{equation} z_1 = \frac{{\boldsymbol{\rho}}_1\cdot{\boldsymbol{\kappa}}_1 - {\boldsymbol{\rho}}_2\cdot{\boldsymbol{\kappa}}_2}{|{\boldsymbol{\rho}}_1|^2+|{\boldsymbol{\rho}}_2|^2},\hspace{0.5cm} z_2 = \frac{{\boldsymbol{\rho}}_1\cdot{\boldsymbol{\kappa}}_2 + {\boldsymbol{\rho}}_2\cdot{\boldsymbol{\kappa}}_1}{|{\boldsymbol{\rho}}_1|^2+|{\boldsymbol{\rho}}_2|^2},\hspace{0.5cm} z_3 = \frac{{\boldsymbol{\rho}}_1\times{\boldsymbol{\kappa}}_2 - {\boldsymbol{\rho}}_2\times{\boldsymbol{\kappa}}_1}{|{\boldsymbol{\rho}}_1|^2+|{\boldsymbol{\rho}}_2|^2}.\end{equation}
Together, the $w_1, w_2, w_3$ and $z_1, z_2, z_3$ form a canonical set of local coordinates of the reduced phase space $T^*\mathcal{S}_R$. On that space, the reduced Hamiltonian dynamics of the $E=\textbf{P}= \textbf{L}=0$ Newtonian universe is formulated.

Geometrically, each vector $\textbf{w}=(w_1, w_2, w_3)$ determines a point on a two-sphere $S^2_{||\textbf{w}||}(0)$ around the origin where \begin{equation}||\textbf{w}||=1/2 I.\end{equation} Here $I=\sum_i m_i \textbf{q}_i^2$ is the moment of inertia of the three-particle system. Hence, while the radius of the two-sphere determines the size of the triangle, the position on the sphere (specified by the two angles $\psi$ and $\phi$) determines its shape. Let the Hopf coordinates $w_1$, $w_2$, and $w_3$ point into the $x$, $y$, and $z$ direction, respectively. In that case, the collinear configurations lie on the equator while the equilateral triangle and its reflected version lie at the top and bottom of the shape sphere.

 \subsection{Central configurations on $T^*\mathcal{S}_R$}
 
To study total collisions, we need to analyze the dynamics at the points of central configuration on shape space (the only points at which a total collision might occur).
Let us specify the five points of central configuration with respect to the Hopf coordinates. Let us, for means of simplicity, consider the equal-mass case: $m_1=m_2=m_3=m$. In that case,
\begin{itemize}

\item the two \textbf{Lagrange configurations} (the equilateral triangle and its reflected version) are specified by
\begin{equation} w_1=w_2=0,\ w_3= \pm ||\textbf{w}||\end{equation}

\item and the three \textbf{Euler configurations} (the three possible collinear configurations where one particle is centered in between the other two) are specified by
\begin{equation} w_3=0\end{equation}
and
\begin{equation}(w_1,w_2) \in \bigg\{\bigg(||\textbf{w}||, 0\bigg), \bigg(- \frac{1}{2}||\textbf{w}||, \frac{\sqrt{3}}{2}||\textbf{w}||\bigg), \bigg(- \frac{1}{2}||\textbf{w}||, - \frac{\sqrt{3}}{2}||\textbf{w}||\bigg)\bigg\}.\end{equation}
\end{itemize}
This is obtained directly from expressing the Lagrange and Euler configurations with respect to the Jacobi and Hopf coordinates defined above. 

We see that the five central configurations determine five points on the two-sphere $S^2_{||\textbf{w}||}$  (where $||\textbf{w}||=1/2I$). Let, again, the Hopf coordinates $w_1$, $w_2$ and $w_3$ point into the $x$, $y$, and $z$ direction, respectively. In that case, the two Lagrange points lie at the top and bottom of the sphere, while the three Euler points lie on the equator at equal distance from each other.

\subsection{Reduced Hamiltonian dynamics on $T^*\mathcal{S}_R$} 

The reduced Hamiltonian formulation of the $E=\textbf{P}= \textbf{L}=0$ Newtonian universe
is due to Barbour et al. [2013]. The reformulation of the Newton potential $V_{N}$ with respect to the Hopf coordinates has been obtained before by Montgomery [2002]. 

With respect to the Hopf coordinates $\textbf{w}$ and their canonical conjugates $\textbf{z}$ the reduced Hamiltonian $H_{0}$ on $T^*\mathcal{S}_R$ can be expressed as follows (cf. Barbour et al. [2013]):
\begin{equation} H = T+V_{N}=||\textbf{w}|| \cdot || \textbf{z} ||^2 + \frac{V_S} {\sqrt{||\textbf{w}||}},\end{equation}
where $V_S=V_S(\textbf{w})$. In accordance with Barbour et al., we call $V_S$ the {`shape potential'}. 
Explicitly, the shape potential is of the following form (cf. Montgomery [2002]):
\begin{equation} V_S = -\sqrt{2} \sum_{i<j} \frac{G(m_im_j)^{\frac{3}{2}}(m_i+m_j)^{-\frac{1}{2}}}{\sqrt{1-\textbf{w}\cdot \textbf{b}_{ij}  /||\textbf{w}||}}. \end{equation}
Here the $\textbf{b}_{ij}$ (with $i<j, i,j=1,2,3$) are the three unit vectors representing the three binary collision points on $\mathcal{S}_R$ (one for each possibility that two of the three particles collide). To be precise, $\textbf{b}_{12}$ represents the collision of particles 1 and 2 (where $|\textbf{q}_{1}-\textbf{q}_2|=0$), and so on. 
Since the $\textbf{b}_{ij}$ are unit vectors, they are uniquely specified by two angles $\psi_{ij}$ and $\phi_{ij}$, that is, they are of the general form 
$\textbf{b}_{ij}=(\sin \psi_{ij} \cos \phi_{ij}, \sin \psi_{ij} \sin \phi_{ij}, \cos \psi_{ij})^T$. Consequently, the product $\textbf{w}\cdot \textbf{b}_{ij}$ is proportional to $||\textbf{w}||$. From this it follows that the shape potential is scale-invariant (this is why it is called the shape potential $V_S$ in the first place).

The physical vector field on $T^*\mathcal{S}_R$ is now determined by the reduced Hamiltonian $H$ from (3.11) according to the standard equations, that is, \begin{equation}\frac{dw_1}{dD}=\frac{\partial H}{\partial z_1}, \hspace{1cm} \frac{dz_1}{dD}= - \frac{\partial H}{\partial w_1},\end{equation} and, analogously, for $w_2$ and $w_3$. 

Inspecting $H$ from (3.11), we find that the kinetic term $T$ is invariant under permutations of $w_1, w_2, w_3$ and $z_1, z_2, z_3$, respectively, but the shape potential $V_S$ is not. This will be important later when we introduce two different choices of spherical coordinates in order to discuss the Euler and Lagrange configurations separately.

Note that due to the $1/\sqrt{||\textbf{w}||}$-dependence of the potential term (see (3.11)), the vector field is still singular at $||\textbf{w}||=0$. This reflects the singularity of the Newton potential at the points of total collision (where $I=0$ and $I\propto ||\textbf{w}||$). We get rid off this singularity only when we describe the system on scale-invariant shape phase space $T^*\mathcal{S}$. Luckily, although the Newtonian dynamics is not scale-invariant, there exists a unique description of the $E=\textbf{P}=\textbf{L}=0$ Newtonian universe on scale-invariant $T^*\mathcal{S}$. This is due to the fact that the dilational momentum $D=\sum \textbf{q}_i \cdot \textbf{p}_i$, the generator of scalings, is monotonic and, hence, can be taken to parametrize the trajectories. 

\subsection{Internal Hamiltonian dynamics on $T^*\mathcal{S}$}

We learned from the Lagrange-Jacobi equation that, within the $E=0$ Newtonian universe, the dilational momentum 
$ D=\sum_i \textbf{q}_i\cdot \textbf{p}_i $ is monotonically increasing along the trajectories (recall that $D= 1/2\dot{I}$ where $\ddot{I}>0$ according to the Lagrange-Jacobi equation (2.4)). This is why we can use $D$ to parametrize the trajectories. 

At the same time, $D$ is the generator of dilations/scalings. This follows directly from the fact that \[D=1/2\dot I\] where $I$ is a measure of the size of the system. That is, a non-zero $D$ generates changes in size. (This is why $D$ is called the dilational momentum in the first place, in direct analogy to linear momentum $\textbf{P}$ and angular momentum $\textbf{L}$, which are the generators of spatial translations and rotations, respectively). 

Note, at this point, that we learn directly from the monotonicity of $D$ that the Newtonian system is not scale-invariant. Since $D$ is monotonically increasing, the quantity $D$ is not conserved, which can be confirmed by directly computing its Poisson bracket with $H$: \begin{equation}\{D, H\}\neq 0.\end{equation} Hence, there is no symmetry of the dynamical law connected to $D$. That is, the Newtonian system is not invariant under scalings/dilations.

However, when we express the motion with respect to the $D$, which we can do because $D$ is monotonic, we get rid of scales precisely because we use the rate of change of scale as a reference parameter. In that description, scale-dependence is eventually captured by the time-dependence ($D$-dependence) of the internal vector field on scale-invariant $T^*\mathcal{S}$. 

Expressing the motion with respect to the internal time parameter $\tau =D$, we will formulate the dynamics on a hypersurface of constant internal time, $\tau=\tau^*$ (respectively, $D=D^*$), just like, on absolute phase space $\Gamma$, motion with respect to absolute time $t$ is formulated on a hypersurface of constant absolute time $t=t^*$. (Originally, the trajectories lie in $\Gamma\times \mathbb{R}$, but in the standard Hamiltonian formulation they are projected onto $\Gamma$.)
To specify a constant-internal-time hypersurface, let without loss of generality $D=0$. 
Let further $H=0$ to take into account that energy is conserved, $E=0$. Note that just like the hypersurface $\Gamma_E= \{(\textbf{q},\textbf{p})\in\Gamma| E(\textbf{q},\textbf{p})=0\}$ is both the space in which the trajectories lie and the space of initial conditions, here the $H=0$, $D=0$ hypersurface (a subset of $T^*\mathcal{S}_R$) is both the space in which the internal, reduced trajectories lie and the space of internal, reduced mid-point data. 

At the same time, since $D$ is the generator of scalings and since $H$ is conjugate to $D$, $\{H,D\}\neq 0$, the $H=0$, $D=0$ hypersurface is a representative of shape phase space $T^*\mathcal{S}$.\footnote{One obtains $T^*S$ from $T^*\mathcal{S}_R$ by setting $D=0$, thus fixing the conserved quantity related to scalings, and setting $H=0$, thus fixing the gauge degree of freedom. Canonically, one would rather fix $I$, which is the canonical conjugate of $D$, instead of ${H}$. However,  since $\{H,D\}\neq 0$, one also obtains a representative of $T^*\mathcal{S}$ by fixing ${H}$.} This is why, when parametrizing the trajectories with respect to $\tau=D$, we obtain a (reduced, internal) Hamiltonian description of the system on shape phase space $T^*\mathcal{S}$.

Let us now express the equations of motion with respect to $\tau= D$. To do that, we need to determine the canonical conjugate of $D$, which, when expressed in terms of the shape variables via the constant-energy constraint $H=0$, will define the {`internal Hamiltonian'} $\mathcal{H}$ governing the motion on $T^*\mathcal{S}$, i.e. the motion with respect to $\tau =D$. 

Expressed in terms of the $\textbf{w}$ and $\textbf{z}$ coordinates from (3.5) and (3.6), the dilational momentum $D$ given in (2.5) becomes \begin{equation}D=  2 \textbf{w}\cdot\textbf{z}.\end{equation} 
From this and the condition of canonical conjugacy, \begin{equation}\{\mathcal{H}, D\}=1,\end{equation} we obtain that
\begin{equation} \mathcal{H}=\log \sqrt{1/2 \cdot I(\textbf{w})}, \end{equation} where $I(\textbf{w}, \textbf{z})$  is the moment of inertia expressed with respect to the $\textbf{w}$ coordinates (3.7): \[ I(\textbf{w})=2||\textbf{w}||.\]
When expressed in terms of the internal, i.e. shape variables by help of the constant-energy constraint $H=0$, $\mathcal{H}$ will be the internal Hamiltonian governing the motion on $T^*\mathcal{S}$ (the $H=D=0$ hypersurface)  with respect to internal time $\tau =D$. In the next section, we will introduce a convenient choice of coordinates on $T^*\mathcal{S}_R$ -- spherical coordinates --, which allows us to disentangle shape and scale (in contrast to the Hopf coordinates where both are mixed). This will enable us to determine the explicit form of $\mathcal{H}$ solely in terms of the shape coordinates, the local coordinates of $T^*\mathcal{S}$.

\section{The points of total collision on shape space}

As mentioned before, in order to obtain the equations of motion on $T^*\mathcal{S}$ from the equations of motion on $T^*\mathcal{S}_R$, we need to separate the shape and scale degrees of freedom, which we do by help of spherical coordinates. 
In what follows, we will introduce two different sets of spherical coordinates in order to discuss the Euler and Lagrange points separately. This is convenient because we can then choose the coordinates such that either the Euler or the Lagrange points lie on the equator of the shape sphere where the reduced, internal vector field is non-singular and of a particularly simple form.

\subsection{Choice of shape coordinates} 

The first choice of spherical coordinates, which allows us to discuss the Euler configurations, is also used by Montgomery [2002] and Barbour, Koslowski, and Mercati ([2013], [2015]). It is such that the collinear configurations (hence, also the Euler configurations) lie on the equator of the shape sphere, while the two Lagrange points are at the top and bottom of the shape sphere. This is the way in which the shape sphere is usually depicted (see, e.g., Barbour et al. [2013]). 

The second choice of spherical coordinates can be interpreted as a rotation of the `Hopfian' coordinate system $w_1, w_2, w_3$ such that $w_1$ becomes $w_2$, $w_2$ becomes $w_3$, and $w_3$ becomes $w_1$ (see below). That way, the two Lagrange configurations are `brought onto' the equator of the shape sphere while the Euler configurations are brought onto a meridian (the intersection of the shape sphere and the `new' $w_3=0$ plane). This is achieved by redefining the two angular coordinates. Note that by redefining the angular coordinates and rotating the whole coordinate system, we can achieve that the coordinates end up in their usual position, which amounts to the before-mentioned rotation of the Hopfian coordinate system.

From one point of view, we use that we can rotate the Hopfian coordinate system the way we like because different orientations of the Hopfian coordinate system don't connect to different physics (we have factored out rotations already). From the other point of view, we simply use that we can define the spherical coordinate system the way we like with respect to the Hopfian coordinate system. Note that both views are just different perspectives on one and the same situation.

\begin{itemize}
\item
\textbf{Euler configurations.} In order to discuss the Euler configurations, let us choose spherical coordinates $R, \psi, \phi$ such that \begin{equation} w_1= R \sin \psi \cos \phi, \hspace{1cm} 
 w_2= R \sin\psi \sin \phi, \hspace{1cm}w_3= R \cos \psi. \end{equation}

Note that, by definition of the spherical coordinates, 
\begin{equation} R=||\textbf{w}||. \end{equation}
Since $||\textbf{w}||=1/2 I$ according to (3.7), it follows that the radius of the shape sphere $R$ is a measure of the size of the system. Now $\psi=\pi/2$ specifies the equator of the shape sphere. Consequently, the Euler configurations lie on the equator of the shape sphere -- since $w_3=0$ from (3.9) holds if and only if $\psi=\pi/2$ -- while the Lagrange configurations lie at the top and bottom of the sphere -- since $w_1=w_2=0$ from (3.8) holds if and only if $\psi=0$ or $\psi=\pi$.

To be precise, the three \textbf{Euler points} are specified by
\begin{equation} (\psi_E, \phi_E) \in \bigg\{\bigg(\frac{\pi}{2},0\bigg), \bigg(\frac{\pi}{2},\frac{2}{3}\pi\bigg), \bigg(\frac{\pi}{2},\frac{4}{3}\pi\bigg)\bigg\}.\end{equation}
This follows directly from (3.9) and (3.10) and the definition of $\psi$ and $\phi$ from above.

Connected to $R, \psi, \phi$, there exist canonical conjugates $p_R, p_\psi, p_\phi$ defined as follows:
 \begin{eqnarray} 
 z_1 &=& \frac{1}{R}( \cos \phi(Rp_R\sin\psi -p_\psi\cos\psi)-p_\phi \sin^{-1}\psi\sin\phi) \nonumber\\
 z_2&=&  \frac{1}{R}( \sin \phi(Rp_R\sin\psi -p_\psi\cos\psi)+p_\phi \sin^{-1}\psi\cos\phi)\nonumber\\
 z_3 &=& - p_R\cos \psi - \frac{1}{R}p_\psi\sin\psi. \end{eqnarray}
 
 Together, $R, \psi, \phi$ and $p_R, p_\psi, p_\phi$ form a canonical basis of $T^*S_R$.

\item \textbf{Lagrange configurations.} In order to discuss the Lagrange configurations, we choose  spherical coordinates $R', \psi', \phi'$ (with $R'=R$) such that 
 \begin{equation} w_1= R' \cos \psi', \hspace{1cm} w_2= R' \sin \psi' \cos \phi',
  \hspace{1cm} w_3= R' \sin\psi' \sin \phi'. \end{equation}
Note that this new choice of coordinates represents a permutation of the $w_1, w_2$, and $w_3$ (a rotation of the axes such that $w_1$ becomes $w_2$ and so on).
Now $\psi'=\pi/2$ denotes the equator of the shape sphere. It follows that now the two Lagrange configurations  lie on the equator of the shape sphere, since $w_1=0$ (3.8) holds if and only if $\psi'=\pi/2$. Now the Euler configurations lie on a meridian, the intersection of the shape sphere and the `new' $w_3=0$ plane. 

To be precise, the two \textbf{Lagrange points} are specified by
\begin{equation} (\psi'_L, \phi'_L) \in \bigg\{\bigg(\frac{\pi}{2},\frac{\pi}{2}\bigg), \bigg(\frac{\pi}{2},\frac{3}{2}\pi \bigg)\bigg\}.\end{equation}
This follows directly from (3.8) and the definition of $\psi'$ and $\phi'$ from above.

Again, connected to $R', \psi', \phi'$, there exist canonical conjugates $p_R', p_\psi', p_\phi'$ defined as follows:
 \begin{eqnarray}  z_1 &=& - p_R'\cos \psi' - \frac{1}{R'}p_\psi'\sin\psi'\nonumber\\
 z_2 &=& \frac{1}{R'}( \cos \phi'(R'p_R'\sin\psi' -p_\psi'\cos\psi')-p_\phi' \sin^{-1}\psi'\sin\phi') \nonumber\\
 z_3&=&  \frac{1}{R'}( \sin \phi'(R'p_R'\sin\psi' -p_\psi'\cos\psi')+p_\phi' \sin^{-1}\psi'\cos\phi'). \end{eqnarray} 
Also $R', \psi', \phi'$ and $p_R', p_\psi', p_\phi'$ form a canonical basis of $T^*\mathcal{S}_R$.
\end{itemize}

Unfortunately, we need two different coordinate system ($R, \psi, \phi$ and $R', \psi', \phi'$, respectively) to treat each the Euler and Lagrange configurations. This is due to the fact that the physical vector field on $T^*\mathcal{S}$ is singular at the top and bottom of the shape sphere (due to a coordinate-singularity at that point, see Appendix A) and, once we put the Euler or the Lagrange configurations on the equator, at least one of the other central configurations lies at the top or bottom of the shape sphere (see the equations (3.8)--(3.10) which determine the Euler and Lagrange points).\footnote{Of course, we might come up with a more complicated definition of spherical coordinates, which allows to treat all the central configurations at once. Given that the vector field is non-singular everywhere except at the top and bottom of the shape sphere, we would need a spherical coordinate system which is tilted with respect to the Hopfian system. But then we wouldn't be able to solve the equations of motion analytically, we wouldn't even be able to write down the shape potential in a nice form, that's why we better keep the two coordinate systems we have introduced above.}

\subsection{Dynamics on $T^*\mathcal{S}_R$ and $T^*\mathcal{S}$ in terms of the new coordinates} 

Since the kinetic term $T$ of the Hamiltonian is invariant under permutations of the $w_1, w_2, w_3$ and $z_1, z_2, z_3$ coordinates, we can write down the physical vector field on $T^*\mathcal{S}_R$ (respectively, the equations of motion) in such a way that it does not distinguish between the two different choices of spherical coordinates (which represent precisely two different permutations of the $\textbf{w}$ and $\textbf{z}$ coordinates). This is possible as long as we don't write down the explicit form of the shape potential $V_S$ (respectively $V_S'$) which is not invariant under the given permutations. 

Expressing $H$ from (3.11) with respect to the unprimed spherical coordinates specified in (4.1) and (4.4), we find that
\begin{equation} H = \frac{p_\psi^2+\sin^{-2}\psi p_\phi^2 +R^2p_R^2}{R} + \frac{V_S(\psi, \phi)} {\sqrt{R}}.\end{equation} 
Analogously, with respect to the primed spherical coordinates from (4.5) and (4.7), $H$ can be written as 
\begin{equation}H = \frac{p_{\psi'}^2+\sin^{-2}\psi' p_{\phi'}^2 +(R')^2(p'_R)^2}{R'} + \frac{V_S'(\psi', \phi')} {\sqrt{R'}}\end{equation}
with $V'_S(\psi', \phi')\neq V_S(\psi, \phi)$. (The explicit expression of $V_S(\psi, \phi)$ and $V'_S(\psi', \phi')$ is given in Appendix B). 

The respective Hamiltonian equations of motion on $T^*\mathcal{S}_R$ are
\begin{eqnarray} \frac{\mathrm{d} \psi}{\mathrm{d} t} &=& \frac{2p_\psi}{R}, \hspace{2.3cm} \frac{\mathrm{d} p_\psi}{\mathrm{d} t} = \frac{2\sin^{-3}\psi\cos\psi p_\phi^2}{R}-\frac{\partial V_S/\partial \psi}{\sqrt{R}},\nonumber\\
\frac{\mathrm{d} \phi}{\mathrm{d} t} &=&\frac{2\sin^{-2}\psi p_\phi}{R}, \hspace{1cm} \frac{\mathrm{d} p_\phi}{\mathrm{d} t} =-\frac{\partial V_S/\partial \phi}{\sqrt{R}},\nonumber\\
\frac{\mathrm{d} R}{\mathrm{d} t} &=&2 R\cdot p_R, \hspace{1.7cm} \frac{\mathrm{d} p_R}{\mathrm{d} t} = \frac{p_\psi^2+\sin^{-2}\psi p_\phi^2- R^2 p_R^2}{R^2} +\frac{1}{2}\frac{V_S(\psi, \phi)}{R^{3/2}}.\end{eqnarray}
and, analogously, for the primed coordinates. Only the functional form of the shape potential depends on the choice of coordinates, i.e., $V_S'(\psi', \phi')\neq V_S(\psi, \phi)$.

Note that, of course, since scale is still part of the description, the equations of motion diverge at $R= 1/2 I=0$ (the point of total collision) representing the singularity of the Newton potential at that point. Since the physical vector field is singular at $R=0$, the solutions cannot be continued through the point of total collision.

However, now that we have separated scale and shape degrees of freedom, we can write down the equations of motion on scale-invariant $T^*\mathcal{S}$ solely in terms of the shape variables (and internal time $\tau =D$). Recall that the equations on $T^*S$ are generated by the internal Hamiltonian $\mathcal{H}$ from (3.17), the canonical conjugate of $\tau=D$. Recall also that $\mathcal{H}$ is expressed on $T^*\mathcal{S}$ (the $H=D=0$ hypersurface) in terms of the internal, i.e. shape variables by help of the constant energy constraint $H=0$. 

Expressed with respect to the spherical coordinates (4.1), 
\begin{equation} \mathcal{H}=  \log \sqrt{1/2 I} =  \log \sqrt{R},\end{equation} 
where we used that $R=||\textbf{w}||=1/2 I$. This becomes a function of the shape coordinates given that $H=0$, that is, $\sqrt{R}\big|_{H=0}= \sqrt{R(\psi, \phi, p_\psi, p_\phi)}$. With $H$ from (4.8), the constant energy condition reads
\begin{equation} H=\frac{p_\psi^2+\sin^{-2}\psi p_\phi^2 +R^2p_R^2}{R} + \frac{V_S(\psi, \phi)} {\sqrt{R}}= 0. \end{equation}
Solving this equation for $R$ and using that, when expressed with respect to the spherical coordinates (4.1) and (4.4), the dilational momentum $D$ from (3.15), namely $D=2\textbf{w} \cdot \textbf{z}$, turns into \begin{equation}D= 2 R\cdot p_R,\end{equation} we obtain:
\begin{equation} \sqrt{R}\big|_{H=0}= \frac{p_\psi^2 + \sin^{-2}\psi p_\phi^2+ 1/4 \tau^2}{-V_S (\psi, \phi)}.\end{equation}
Here we have set $\tau=D$.
Inserting this into (4.11), we obtain the internal Hamiltonian $\mathcal{H}$ governing the motion on $T^*\mathcal{S}$ with respect to internal time $\tau$:
\begin{equation} \mathcal{H}= \log \frac{p_\psi^2 + \sin^{-2}\psi p_\phi^2+ 1/4 \tau^2}{-V_S (\psi, \phi)}\end{equation}
Note that this Hamiltonian is time-dependent ($\tau$-dependent). This is the last remnant of scale, incorporating the scale-dependence of the system, in the otherwise scale-free description on shape phase space $T^*S$.

We can now write down the Hamiltonian equations of motion on $T^*\mathcal{S}$, $\mathrm{d}\psi/\mathrm{d}\tau = \partial \mathcal{H}/\partial p_\psi$,  $\mathrm{d}p_\psi/\mathrm{d}\tau = -\partial \mathcal{H}/\partial \psi$, and so on:
\begin{eqnarray} \frac{\mathrm{d}\psi}{\mathrm{d}\tau} &=&  \frac{2p_{\psi}}{p_{\psi}^2+\sin^{-2}\psi p_{\phi}^2 +\frac{1}{4}\tau^2}, \hspace{1cm}
\frac{\mathrm{d}p_{\psi}}{\mathrm{d}\tau} = \frac{2\sin^{-3}\psi\cos\psi p_{\phi}^2}{p_{\psi}^2+\sin^{-2}\psi p_{\phi}^2 +\frac{1}{4}\tau^2}+\frac{\partial \log (-V_S)}{\partial\psi}, \nonumber\\
 \frac{\mathrm{d}\phi}{\mathrm{d}\tau}&=& \frac{2\sin^{-2}\psi p_{\phi}}{p_{\psi}^2+\sin^{-2}\psi p_{\phi}^2 +\frac{1}{4}\tau^2},\hspace{1cm}
 \frac{\mathrm{d}p_{\phi}}{\mathrm{d}\tau}=\frac{\partial \log (-V_S)}{\partial\phi}.
 \end{eqnarray}
Again, for the primed variables, the equations of motion are analogous (only that $\psi=\psi'$, $\phi=\phi'$, $p_\psi=p'_\psi, p_\phi=p'_\phi$, and $V_S=V_S'$).

\subsection{Equations of motion at the points of total collision}
 
In what follows, we want to derive the equations of motion on shape phase space $T^*S$ at the Janus point  ($\tau=0$) at the central configurations. 

Remember that we chose spherical coordinates such that, for each choice, the central configurations under consideration lie on the equator of the shape sphere. In other words, it suffices to analyze the vector field (4.16) at the equator, that is, at $\psi =\pi/2$ ($\psi'=\pi/2$). Since $\sin \pi/2=1$ and $\cos \pi/2=0$, we obtain the equations
 \begin{eqnarray} \frac{\mathrm{d}\psi}{\mathrm{d}\tau} &=&  \frac{2p_{\psi}}{p_{\psi}^2+ p_{\phi}^2+\frac{1}{4}\tau^2}, \hspace{1cm}
\frac{\mathrm{d}p_{\psi}}{\mathrm{d}\tau} =\bigg[\frac{\partial \log (-V_S)}{\partial\psi}\bigg]_{\psi=\frac{\pi}{2}}, \nonumber\\
 \frac{\mathrm{d}\phi}{\mathrm{d}\tau}&=& \frac{2 p_{\phi}}{p_{\psi}^2+ p_{\phi}^2+\frac{1}{4}\tau^2},\hspace{1cm}
 \frac{\mathrm{d}p_{\phi}}{\mathrm{d}\tau}=\bigg[ \frac{\partial \log (-V_S)}{\partial\phi}\bigg]_{\psi=\frac{\pi}{2}} 
 \end{eqnarray}
and, analogously, for the primed coordinates.
 
To evaluate this vector field at the central configurations $(\psi_{E}, \phi_{E})$ and $(\psi'_{L}, \phi'_{L})$, it remains to determine the partial derivatives $\partial \log (-V_S)/\partial\psi$ and $\partial \log (-V_S)/\partial\phi$ (respectively, $\partial \log (-V'_S)/\partial\psi'$ and $\partial \log (-V'_S)/\partial\phi'$) at these points. Since, for the two different choices of spherical coordinates, the functional form of the shape potential differs, we have to analyze each the Euler and Lagrange configurations separately. The respective computation of the partial derivatives of the logarithm of the shape potential is given in Appendix B. 

In Appendix B, it is shown that all partial derivatives vanish identically (see (5.3) -- (5.6)). Consequently, for the Euler configurations $(\psi_E, \phi_E)$,
\begin{equation} \bigg[\frac{\partial \log (-V_S)}{\partial\psi}\bigg]_{(\psi_E, \phi_E)}=0,\hspace{1cm} 
 \bigg[ \frac{\partial \log (-V_S)}{\partial\phi}\bigg]_{(\psi_E, \phi_E)}=0.
 \end{equation}
 and for the Lagrange configurations $(\psi'_L, \phi'_L)$, 
 \begin{equation} \bigg[\frac{\partial \log (-V'_S)}{\partial\psi'}\bigg]_{(\psi'_L, \phi'_L)}=0,\hspace{1cm} 
 \bigg[ \frac{\partial \log (-V'_S)}{\partial\phi'}\bigg]_{(\psi'_L, \phi_L')}=0.
 \end{equation}

The fact that, at the Euler and Lagrange points, all partial derivatives of $\log(-V_S)$ vanish tells us that the central configurations are stationary points of the (negative) shape potential. This follows from the fact that $\partial \log (-V_S)/ \partial x = V_S^{-1} \partial V_S/\partial x$ and $0<V_S(\psi_{E}, \phi_{E})<\infty$ ($0<V'_S(\psi'_L, \phi'_L)<\infty$). Hence, at the central configurations, $\partial \log (-V_S)/ \partial \psi=0$ if and only if $\partial V_S/\partial \psi =0$ (and, analogously, for $\partial V_S/\partial \phi$). Since all partial derivatives are zero, the central configurations are stationary points of the shape potential.\footnote{Being stationary points, each of the central configurations is either a minimum or a maximum or a saddle point of the shape potential. If we look at the second derivatives or, more precisely, at the eigenvalues of the Hessian matrix with entries $\partial^2 V_S/\partial^2 \psi$, $\partial^2 V_S/\partial \psi \partial \phi$, $\partial^2 V_S/\partial \phi \partial \psi$ and $\partial^2 V_S/\partial^2\phi$, we find that the Lagrange configurations are maxima, while the Euler configurations are saddle points of the shape potential.}

Inserting (4.18) and (4.19) into (4.17) we find that, both at the Euler and Lagrange configurations, the equations of motion are of the following form: 
 \begin{eqnarray} \frac{\mathrm{d}\psi}{\mathrm{d}\tau} &=&  \frac{2p_{\psi}}{p_{\psi}^2+ p_{\phi}^2+\frac{1}{4}\tau^2}, \hspace{1cm}
\frac{\mathrm{d}p_{\psi}}{\mathrm{d}\tau} =0, \nonumber\\
 \frac{\mathrm{d}\phi}{\mathrm{d}\tau}&=& \frac{2 p_{\phi}}{p_{\psi}^2+ p_{\phi}^2+\frac{1}{4}\tau^2},\hspace{1cm}
 \frac{\mathrm{d}p_{\phi}}{\mathrm{d}\tau}=0.
 \end{eqnarray}

In order to study total collisions, we have to evaluate these equations at $\tau=0$ (where $I=I_{min}$, the only moment at which a total collision may occur). At $\tau=0$ at the Euler and Lagrange points, the vector field on $T^*S$ is 
 \begin{eqnarray} \frac{\mathrm{d}\psi}{\mathrm{d}\tau} &=&  \frac{2p_{\psi}}{p_{\psi}^2+ p_{\phi}^2}, \hspace{1cm}
\frac{\mathrm{d}p_{\psi}}{\mathrm{d}\tau} =0, \nonumber\\
 \frac{\mathrm{d}\phi}{\mathrm{d}\tau}&=& \frac{2 p_{\phi}}{p_{\psi}^2+ p_{\phi}^2},\hspace{1cm}
 \frac{\mathrm{d}p_{\phi}}{\mathrm{d}\tau}=0.
 \end{eqnarray}
This vector-field is regular for almost all solutions. Unfortunately, it is singular precisely on the set of total collisions!

Regarding the first assertion, note that the vector field (4.21) is non-singular and thus well-defined for all $p_\psi$ and $p_\phi$ except if $p_\psi=p_\phi=0$. Recall, at this point, that every solution passing a central configuration (Euler or Lagrange point) at $\tau =0$ is uniquely specified by mid-point data $(\psi, \phi, p_\psi, p_\phi)(0)= (\psi_{E/L}, \phi_{E/L}, p_\psi(0), p_\phi(0))$. Among these mid-point data, those with shape momenta $p_\psi(0)=p_\phi(0)=0$ form a set of measure zero (see section 4.4).

Still, this measure-zero set is precisely the set of solutions which feature a total collision (see section 4.5). All mid-point shape momenta $p_\psi(0)\neq 0$ and $p_\phi(0)\neq 0$ (where it suffices that one of the shape momenta is non-zero) determine solutions of the form that the system passes a central configuration at the Janus point, $\tau =0$, with some non-zero minimal moment of inertia, $I_{min}\neq 0$.

\subsection{Measure-zero set of solutions}

To see that the set of solutions with $p_\psi(0)=p_\phi(0)=0$ forms a set of measure zero among all solutions, consider the canonical volume measure (the reduced Liouville measure) $\mu$ on $T^*S$,
 \begin{equation} \mu =\mathrm{d}\psi\mathrm{d}\phi\mathrm{d}p_\psi\mathrm{d}p_\phi.\end{equation}
This measure is derived in Appendix C. It is obtained from the original Liouville measure $\prod_{i=1}^{3N} \mathrm{d}{q}_i\mathrm{d}{p}_i$ on $\Gamma$ by fixing the momentum constraints and factoring out the gauge volume. It is stationary, that is, conserved under the internal Hamiltonian time evolution. Hence, it is the appropriate measure for the statistical analysis of the system.

Let, in what follows, $\Gamma_{E,L}\subset T^*\mathcal{S}$ denote the set of solutions which, at $\tau=0$, pass a central configuration,
\[\Gamma_{E,L}=\{(\psi,\phi,p_\psi,p_\phi)\in T^*\mathcal{S}|(\psi, \phi)=(\psi_{E/L}, \phi_{E/L})\},\]
and let $\Gamma_S\subset \Gamma_{E,L}$ further denote the subset of points where the vector field (4.21) is singular, i.e., 
\[\Gamma_S=\{(\psi,\phi,p_\psi,p_\phi)\in T^*\mathcal{S}|(\psi, \phi)=(\psi_{E/L}, \phi_{E/L}),p_\psi=p_\phi=0\}.\]
In what follows, let us determine the conditional measure $\mu(\Gamma_S| \Gamma_{E,L})$ of the set of solutions which, at $\tau=0$, pass a central configuration with $p_\psi=p_\phi=0$. According to the measure $\mu$ on $T^*S$,
\begin{eqnarray} \mu(\Gamma_S| \Gamma_{E,L})&=& \frac{\mu(\Gamma_S)}{\mu(\Gamma_{E,L})} = \frac{ \int_{T^*\mathcal{S}} \delta(\psi-\psi_{E/L})\delta(\phi-\phi_{E,L})\delta(p_\psi)\delta(p_\phi){d}\psi{d}\phi{d}p_\psi{d}p_\phi }{\int_{T^*\mathcal{S}} \delta(\psi-\psi_{E/L})\delta(\phi-\phi_{E,L})\mathrm{d}\psi\mathrm{d}\phi\mathrm{d}p_\psi\mathrm{d}p_\phi}\nonumber\\
&=&\frac{1}{\int_{\mathbb{R}^2}\mathrm{d}p_\psi\mathrm{d}p_\phi }=0.\end{eqnarray}

We have just shown that, among all solutions which pass a central configuration at $\tau=0$, those with $p_\psi=p_\phi=0$ (which are precisely those which pass a total collision, see section 4.5) form a set of measure zero. Even more, since $\Gamma_{E/L}\subset T^*S$ (and $T^*S\subset \Gamma$), they form a set of measure zero on all of $T^*S$ (and $\Gamma$), that is, among all solutions.

\subsection{Shape-dynamical description of total collisions}

In order to see why, at the moment of total collision, the shape momenta are necessarily zero, we have to go back to the constant-energy condition (4.12). Using the relation $\tau=2R\cdot p_R$ (see (4.13)), (4.12) can be rewritten as
\[ H=\frac{p_\psi^2+\sin^{-2}\psi p_\phi^2 +1/4 \tau^2}{R} + \frac{V_S(\psi, \phi)} {\sqrt{R}}= 0. \] 
If we multiply this constraint with $R$, we obtain
\begin{equation} {p_\psi^2+\sin^{-2}\psi p_\phi^2 +1/4 \tau^2} + V_S(\psi, \phi) \sqrt{R}= 0. \end{equation}

Now study this constraint in the limit $R\to 0$ and $\psi \to \psi_{E,L}$, $\phi \to \phi_{E, L}$ (the limit of a total collision). 
We know that, at the central configurations, $\psi_{E/L}=\pi/2$ (hence, $\sin \psi_{E/L}=1$). Moreover, $|V_S(\psi_{E/L}, \phi_{E/L})|<\infty$ (to see this, insert $(\psi_{E/L}, \phi_{E/L})$ into (5.1), respectively (5.2), or note that the shape potential diverges only at the points of binary collision). It follows that, in the limit $\psi \to \psi_{E,L}$, $\phi \to \phi_{E, L}$, the constraint equation turns into
\begin{equation} {p_\psi^2+ p_\phi^2 +1/4 \tau^2} + V_S(\psi_{E/L}, \phi_{E/L}) \sqrt{R}= 0,\end{equation}
where $|V_S(\psi_{E/L}, \phi_{E/L})| <\infty$.

In the limit $R\to 0$, (4.25) holds if and only if, in that limit, $p_\psi$, $p_\phi$ and $\tau$ converge to zero. In other words, in case a total collision occurs, necessarily $\tau =0$ (which we knew) and $p_\psi=p_\phi=0$. The other way round, if $\tau=0$ and $p_\psi=p_\phi=0$, it follows from (4.25) that the particles have to collide, $R=0$.

Note that we get this result almost alone from the limit $R\to 0$, that is, without demanding that the solutions converge to a central configuration. All we need  is that $|V_S(\psi, \phi)|<\infty$ (respectively, $|V_S'(\psi', \phi')|<\infty$). If the shape potential is bounded, (4.24) holds in the limit $R\to 0$ if and only if, in that limit, $\tau\to0$ and $p_\psi\to 0$, $p_\phi\to 0$.\footnote{To obtain this stronger result, we would have to deal with the binary collision points $(\psi_{ij}, \phi_{ij})$, the (only) points where the shape potential diverges (inspect (5.1) and (5.2)), separately. Unfortunately, we cannot exclude the binary-collision points \textit{a priori}. Assume, for instance, that two particles collide just as fast as $R\to 0$, while the third particle collides with the two of them only as fast as $\sqrt{R}\to 0$. In that case, a total collision occurs precisely at a binary collision point. Then $-V_S\sim 1/\sqrt{R}$ and the last term in (4.24) would become a constant in the limit $R\to 0$. Since that term is negative ($V_S$ is negative), it could be encountered by a positive term, a sum of squares of non-zero $p_\psi, p_\phi$ and $\tau$. In that case, we would no longer be able to conclude that $p_\psi=p_\phi=\tau=0$ in the limit $R \to 0$. Hence, to obtain the stronger result, namely that $R=0$ if and only if $p_\psi =p_\phi =\tau=0$ (without assuming the particles to form a central configuration), we would have to show that a behavior as the one described above, with a total collision at a binary collision point, cannot happen.}

We have thus obtained a purely shape-dynamical description of total collisions. Recall that in (2.8) we defined a total collision via the vanishing moment of inertia: \[I=2R=0.\] Now we are able to define it merely in terms of $\tau$ and the shape variables, that is, without reference to external scale. According to this new definition, a total collision occurs if and only if, at the moment of minimal extension, $\tau=0$, the particles form a central configuration and all shape momenta are zero, that is, if and only if, at $\tau=0$,
\begin{equation} (\psi, \phi)=(\psi_{E/L}, \phi_{E/L}) \hspace{0.2cm} \mathrm{and} \hspace{0.2cm} p_\psi = p_\phi=0. \end{equation}
Importantly, this definition is free of external scale. It is a definition in purely relational, i.e. shape-dynamical, terms.

\section{Conclusion}

This paper has shown that, even though a purely relational formulation of the $E=\textbf{P}=\textbf{L}=0$ Newtonian universe of three bodies exists -- a formulation, that is, of the Newtonian gravitational system on shape phase space -- and even though this implies that absolute scale is no longer part of the (unique) description of the system, the Newtonian singularity at the points of total collision remains. 

The relational analysis has shown that, at the points of total collision, the equations of motion are singular due to the particular behavior of the shape momenta, which are necessarily zero at these points. While this  implies that the singularity remains, it also has one positive effect. It allows us to define total collisions in purely shape-dynamical terms, that is, without reference to (external) absolute scale. Instead, we are able to distinguish total-collision solutions from non-collision solutions on shape phase space. This shape-dynamical description we also use to show that solutions featuring a total collision form a set of measure zero among all solutions. 

The general form of the analysis suggests that all results, which have been obtained for the three-body model, should hold for the $N$-body model as well. That is, also for $N$ particles, we expect the shape equations of motion to be singular at the points at (and only at) which the shape momenta are all zero. And again, total collisions on $6N-14$ dimensional shape phase space will be points of central configuration with vanishing shape momenta at the moment of minimal extension of the system. Of course, for $N$ particles, we won't be able to make the explicit computations, but instead will have to give arguments of a more general and abstract form. But this is merely a technical issue and should be possible in principle. 

Although the result is negative in the sense that, even on shape space, the singularity remains, the original  idea -- to glue together two solutions from absolute space via the unique evolution of certain relational (that is, shape) variables -- need not be dismissed altogether. Although the shape variables we presented didn't serve for that purpose, one might still hope to find some other variables, singular variables, which allow for an evolution though the points of total collision. But this will involve a much harder study of the singularity, a topic for future research.

\newpage

\noindent \Large \textbf{Appendix}\\

\large \noindent \textbf{Appendix A: Singularity of the physical vector field at $\psi=0$} \\

\normalsize
\noindent We show that the vector field (4.16) is singular at $\psi=0$ and $\psi=\pi$ (top and bottom of the shape sphere). An analogous result holds for the primed coordinates (at $\psi'=0$ and $\psi'=\pi$).

Consider the equations of motion on $T^*\mathcal{S}$ (4.16). The corresponding physical vector field is singular at $\psi=0$ and $\psi=\pi$ if and only if at least one of the equations diverges.
 
Consider the equation for $p_\psi$,
\[\frac{\mathrm{d}p_{\psi}}{\mathrm{d}\tau} = \frac{2\sin^{-3}\psi\cos\psi p_{\phi}^2}{p_{\psi}^2+\sin^{-2}\psi p_{\phi}^2 +\frac{1}{4}\tau^2}+\frac{\partial \log (-V_S)}{\partial\psi}.\]
In the limit $\psi\to 0$ and $\psi\to\pi$, it is 
\begin{eqnarray} \lim_{\psi\to 0/\pi}\bigg[\frac{2\sin^{-3}\psi\cos\psi p_{\phi}^2}{p_{\psi}^2+\sin^{-2}\psi p_{\phi}^2 +\frac{1}{4}\tau^2}\bigg] &=& \lim_{\psi\to 0/\pi}\bigg[ \frac{-6\sin^{-4}\psi\cos\psi p_\phi^2 -2\sin^{-2}\psi p_\phi^2}{-2\sin^{-3}\psi p_\phi^2}\bigg] \nonumber\\
&=& \lim_{\psi\to 0/\pi}\bigg[ \frac{3\cos\psi}{\sin\psi} +\sin\psi \bigg]=\infty.\nonumber\end{eqnarray}
This shows the assertion.

Note that this singularity is not a physical singularity, but a coordinate singularity. It can also be read off the Jacobian, that is, the determinant of the transformation matrix from the $\textbf{w}$ to the spherical coordinates,
\[ \det \frac{d(w_1, w_2, w_3)}{{d}(R, \psi, \phi)} = R^2 \sin\psi,\]
which is zero at $R=0$, $\psi=0$, and $\psi=\pi$. \\

\large
\noindent \textbf{Appendix B: Partial derivatives of the shape potential $V_{S}$} \\

\normalsize
\noindent From (3.12) we know that the shape potential $V_S=V_S(\textbf{w})$ can be written as
\[ V_S = -\sqrt{2} \sum_{i<j} \frac{G(m_im_j)^{\frac{3}{2}}(m_i+m_j)^{-\frac{1}{2}}}{\sqrt{1-\textbf{w}\cdot \textbf{b}_{ij}  /||\textbf{w}||}}. \]
Recall that the $\textbf{b}_{ij}$ are the unit vectors which represent the three binary collisions (specified by two angles $\psi_{ij}$ and $\phi_{ij}$ and where the $i$ and $j$ refer to the collision particles). Let us determine the form of $V_S$ for the two different choices of spherical coordinates.

\begin{itemize}
\item \textbf{Euler configurations.} The binary collision vectors always lie on the $w_3=0$ plane. For the unprimed coordinates (4.1) and (4.4), this means that they lie on the equator of the shape sphere. The equator is specified by the angle $\psi=\pi/2$, hence, $\psi_{ij}=\pi/2$. If the particles have equal masses, the three binary collision points are further specified by $\phi_{ij}\in\{\frac{1}{3}\pi, \pi,\frac{5}{3}\pi\}$ (where, again, the $i$ and $j$ refer to the collision particles). Hence, for the given coordinates,
\[ \textbf{w}\cdot \textbf{b}_{ij}  = ||\textbf{w}||\sin \psi \cos (\phi-\phi_{ij}).\]
Here the $(\phi-\phi_{ij})$ are the angles between the $\textbf{b}_{ij}$ and the projection of $\textbf{w}$ onto the $w_3=0$ plane and the term $||\textbf{w}||\sin \psi$ is the component of $\textbf{w}$ which is parallel to the $w_3=0$ plane.

The shape potential can then be written as
\begin{equation} V_S = - \sqrt{2} \sum_{i<j} \frac{G(m_im_j)^{\frac{3}{2}}(m_i+m_j)^{-\frac{1}{2}}}{\sqrt{1-\sin \psi \cos (\phi-\phi_{ij})}}.\end{equation}
This is the form of the shape potential for the choice of coordinates which is appropriate to discuss the Euler configurations.

\item \textbf{Lagrange configurations.} Given the primed coordinates (4.5) and (4.7), the $w_3=0$ plane is specified by $\phi'_{ij}=0$, respectively $\phi'_{ij}=\pi$. (In this case, the intersection of the shape sphere and the $w_3=0$ plane is a meridian, not the equator.) The three binary collision vectors are further specified by three angles $\psi'_{ij}$ (with $i<j$ and $i,j=1,2,3$). In the equal-mass case, the respective collision vectors are $\textbf{b}_{ij}=(\phi'_{ij}, \psi'_{ij})\in \{(0, \frac{1}{3}\pi), (0, \pi), (\pi, \frac{1}{3}\pi)\}$. 
We now have 
\[ \textbf{w}\cdot \textbf{b}_{ij}  =  ||\textbf{w}|| |\cos \phi' | \cos(\psi'\pm \psi'_{ij}). \]
In this setting, $||\textbf{w}|| \cdot |\cos \phi'|$ is the component of $\textbf{w}$ parallel to the $w_3=0$ plane and $(\psi'\pm \psi'_{ij})$ is the angle between the projection of $\textbf{w}$ onto the $w_3=0$ plane and the binary collision vector (where the plusminus takes into account that $\psi$ runs from 0 to $\pi$ and not from 0 to $2\pi$, hence, in order to correctly measure the angle between the vectors $\textbf{w}$ and $\textbf{b}_{ij}$ we need to adjust the sign appropriately, where we have to take `$-$'  if both vectors are on the same side of the $w_2=0$ plane and `$+$' in case they are on opposite sites).

The shape potential can now be written as
\begin{equation} V'_S = - \sqrt{2} \sum_{i<j} \frac{G(m_im_j)^{\frac{3}{2}}(m_i+m_j)^{-\frac{1}{2}}}{\sqrt{1- |\cos \phi'|  \cos (\psi'\pm\psi'_{ij})}}.\end{equation}
This is the form of the shape potential for the choice of coordinates which is appropriate to discuss the Lagrange configurations.
\end{itemize}

Let us now determine the partial derivatives of $\log (-V_S)$ and $\log (-V'_S)$ at the respective central configurations. For means of simplicity, let us, in what follows, consider the equal mass case, i.e., $m_i=m $ with $i=1, 2, 3$. Then $(m_im_j)^{\frac{3}{2}}(m_i+m_j)^{-\frac{1}{2}}=m^{5/2}/\sqrt{2}$.

\begin{itemize}

\item \textbf{Euler configurations.} Given the first choice of coordinates, the three Euler configurations are specified by $(\psi_E, \phi_E)\in\{(\frac{\pi}{2},0), (\frac{\pi}{2},\frac{2}{3}\pi), (\frac{\pi}{2},\frac{4}{3}\pi)\}$. This is again the equal mass case. Recall that, in that case, the binary collision vectors were specified by $\psi_{ij} =\pi/2$ (which we don't need here since $\psi_{ij}$ doesn't appear in the $V_S$) and $\phi_{ij}\in \{\frac{1}{3}\pi, \pi, \frac{5}{3}\pi\}$. Then, with $V_S$ from (5.1) and given $(\psi_E, \phi_E)$, we obtain:
 \begin{equation} \bigg[\frac{\partial \log (-V_S)}{\partial\psi}\bigg]_{\psi_E, \phi_E} =\bigg[ \frac{Gm^{5/2}}{{2}} \sum_{i<j} \frac{ \cos \psi \cos (\phi-\phi_{ij})}{\sqrt{1- \sin \psi\cos (\phi-\phi_{ij})}^3}\bigg]_{\psi_E, \phi_E} = 0.\end{equation}
This follows from $\cos \psi_E=\cos \pi/2=0$. On the other hand, inserting $\sin \psi_E=\sin \pi/2=1$ and using the fact that, for every $\phi_E$, $(\phi_E-\phi_{ij})\in\{0, \pm1/3\pi, \pm5/3\pi\}$, we obtain that:  
 \begin{eqnarray} \bigg[\frac{\partial \log (-V_S)}{\partial\phi}\bigg]_{\psi_E, \phi_E} &=& \bigg[ - \frac{Gm^{5/2}}{{2}} \sum_{i<j} \frac{\sin \psi \sin(\phi-\phi_{ij})}{\sqrt{1- \sin \psi\cos (\phi-\phi_{ij})}^3} \bigg]_{\psi_E, \phi_E} \nonumber\\
&=& - \frac{Gm^{5/2}}{{2}} \bigg[ \sum_{i<j} \frac{ \sin(\phi-\phi_{ij})}{\sqrt{1- \cos (\phi-\phi_{ij})}^3}\bigg]_{\phi_E} =\nonumber\\ 
&=& -  \frac{Gm^{5/2}}{{2}} \bigg[  \frac{ \sin(-1/3\pi)}{\sqrt{1-\cos(-1/3\pi)}^3} +  0+  \frac{ \sin(-5/3\pi)}{\sqrt{1-\cos(-5/3\pi)}^3}\bigg]\nonumber\\
&=& 0. \end{eqnarray}

To obtain the last equation, we use the (anti-)symmetry relations $\sin (1/3\pi)=-\sin(5/3\pi)$ and $\cos (1/3\pi) =\cos(5/3)\pi$.

\item \textbf{Lagrange configurations.} Given the second choice of coordinates, the two Lagrange configurations are specified by $(\psi'_L,\phi'_L)\in\{ (\frac{\pi}{2},\frac{\pi}{2}), (\frac{\pi}{2},\frac{3}{2}\pi)\}$. In this case, remember that the binary collision vectors were specified by $(\psi'_{ij}, \phi'_{ij})\in \{(\frac{1}{3}\pi, 0), (\pi, 0), (\frac{1}{3}\pi, \pi)\}$. Then, with $V'_S$ given by (5.2), we have:
 \begin{equation} \bigg[\frac{\partial \log (-V'_S)}{\partial\psi'}\bigg]_{\psi'_L, \phi'_L} =\bigg[ - \frac{Gm^{5/2}}{{2}} \sum_{i<j} \frac{|\cos \phi' |\sin (\psi'\pm\psi'_{ij})}{\sqrt{1- |\cos \phi'|\cos (\psi'\pm \psi'_{ij})}^3}\bigg]_{\psi'_L, \phi'_L} = 0.\end{equation}
 This follows from inserting $\phi_L'\in\{\frac{\pi}{2},\frac{3}{2}\pi\}$ into the cosine: $\cos \phi_L'=0$. On the other hand, since $\cos \phi_L'=0$ and $|\sin \phi_L'|= 1$, we obtain  
 \begin{eqnarray} \bigg[\frac{\partial \log (-V'_S)}{\partial\phi'}\bigg]_{\psi'_L, \phi'_L} &=&\bigg[- \frac{1}{{2}} \sum_{i<j} \frac{-|\sin\phi'| \cos(\psi'\pm\psi'_{ij})}{\sqrt{1- |\cos \phi' |\cos (\psi'\pm\psi'_{ij})}^3} \bigg]_{\psi'_L, \phi'_L} \nonumber\\
&=& \frac{Gm^{5/2}}{{2}}\bigg[\sum_{i<j}  \cos(\psi'\pm\psi'_{ij})\bigg]_{\psi'_L}=\nonumber\\
&=& \frac{Gm^{5/2}}{{2}}\bigg[\sum_{i<j}  (\cos(1/6 \pi) + \cos (\pi/2) + \cos (5/6\pi)) \bigg]=  0.  \end{eqnarray}
\end{itemize}
To obtain the last equation, we used that $\cos (1/6\pi)=-\cos(5/6\pi)$ and $\cos (\pi/2) =0$.

\newpage
\large \noindent \textbf{Appendix C: Measure-zero set of solutions}\\

\normalsize

\noindent Since the internal vector field of the $E=\textbf{P}=\textbf{L}=0$ Newtonian universe on $T^*\mathcal{S}$ is a Hamiltonian vector field, it follows that shape phase space volume is conserved under the internal time evolution. In other words, the volume measure on $T^*\mathcal{S}$ is stationary.

To obtain the volume measure on $T^*S$, we use a formula of Faddeev [1969] for the reduced volume measure/the Liouville measure on the reduced phase space. Faddeev shows that, if the reduced phase space is obtained from some original phase space $\Gamma$ by imposing constraints $H_a=0$, $\chi_b=0$ with $ \{H_a,\chi_b\}\neq 0$, then the volume measure (Liouville measure) on the reduced phase space is 
\begin{equation} \mu=  |\det \{H_a, \chi_b\}|\prod_{a,b}\prod_i \delta(H_a)\delta(\chi_b)\mathrm{d}q_i \mathrm{d}p_i.\end{equation}
Here $q_i, p_i$ are the canonical coordinates on $\Gamma$.

In our case, $\Gamma \cong \mathbb{R}^{18}$ with canonical coordinates $\textbf{q}_i, \textbf{p}_i\in\mathbb{R}^3$ and the pairs of conjugate constraints are $\{\textbf{P},\textbf{Q}_{cm}\}, \{\textbf{L},\textbf{I}_L\},$ and $\{D,H\}$ (see section 3.1).
Inserting this in (5.7), we find that the reduced volume measure $\mathrm{d}\mu$ on $T^*S$ is given by
\begin{equation} \mu=|\det\{(\textbf{P},\textbf{L}, D),(\textbf{Q}_{cm},\textbf{I}_L, H)\}|  \prod_{i=1}^3  \delta(\textbf{P}) \delta(\textbf{L}) \delta(D) \delta(\textbf{Q}_{cm}) \delta(\textbf{I}_L)\delta(H)\mathrm{d}\textbf{q}_i \mathrm{d}\textbf{p}_i.\end{equation}

To abbreviate the notation, let us define $A:=\{(\textbf{P},\textbf{L}, D),(\textbf{Q}_{cm},\textbf{I}_L, H)\}$. Explicitly, $A$ is the matrix 
\[ A= \begin{pmatrix} \{\textbf{P}, \textbf{Q}_{cm}\} & \{\textbf{L}, \textbf{Q}_{cm}\}  & \{D, \textbf{Q}_{cm}\}  \\
\{\textbf{P}, \textbf{I}_L\}  & \{\textbf{L}, \textbf{I}_{L}\}  & \{D, \textbf{I}_{L}\} \\
\{\textbf{P},H\} & \{\textbf{L}, H\} & \{D, H\} \end{pmatrix}. \]

Now we can use that $\textbf{P}$ and $\textbf{L}$ are conserved quantities of motion, i.e.,
$ \{\textbf{P}, H\}=\{\textbf{L}, H\}=0.$
Using this, it is easy to compute the determinant of $A$:
\begin{equation} |\det A|= |\det\{(\textbf{P},\textbf{L}, D),(\textbf{Q}_{cm},\textbf{I}_L,H)\}|= |\{\textbf{P}, \textbf{Q}_{cm}\} \{\textbf{L}, \textbf{I}_L\}\{D,H\}|=|\{D,H\}|,\end{equation}
where the last equation follows from the fact that $\{\textbf{P},\textbf{Q}_{cm}\}=\{\textbf{L}, \textbf{I}_L\}=1.$

Inserting (5.9) in (5.8) and using the definition of the Jacobi and Hopf coordinates from section (3.2) as well as the definition of the spherical coordinates (where everything is analogous both in the primed and unprimed coordinates), we find that
\begin{eqnarray}  &&\int |\det\{(\textbf{P},\textbf{L}, D),(\textbf{Q}_{cm},\textbf{I}_L,H)\}|\prod_{i=1}^3 \delta(D)\delta(H) \delta(\textbf{P}) \delta(\textbf{Q}_{cm})  \delta(\textbf{L})\delta(\textbf{I}_L)  \mathrm{d}\textbf{q}_i \mathrm{d}\textbf{p}_i\nonumber\\
&=&\int |\{D,H\}| \delta(D)\delta(H) \delta(\textbf{P})  \delta(\textbf{Q}_{cm}) \delta(\textbf{L})\delta(\textbf{I}_L)\mathrm{d}\textbf{P}\mathrm{d}\textbf{Q}_{cm}\mathrm{d}\textbf{L}\mathrm{d}\textbf{I}_L\mathrm{d}R \mathrm{d}\psi \mathrm{d}\phi  \mathrm{d}p_R \mathrm{d}p_\psi  \mathrm{d}p_\phi\nonumber\\
&=&\int |\{D,H\}^*| \delta(D)\delta(H)\mathrm{d}R \mathrm{d}\psi \mathrm{d}\phi  \mathrm{d}p_R \mathrm{d}p_\psi  \mathrm{d}p_\phi.\end{eqnarray}
Here $\{D,H\}^*$ denotes the Poisson bracket on $T^*\mathcal{S}_R$, i.e., the Poisson bracket evaluated on the constraint surface determined by $\textbf{P}= \textbf{L}=\textbf{Q}_{cm}=\textbf{I}_L=0$. To compute the integral (5.10), we use that
\[ \delta^{(n)}(\textbf{f}(\textbf{x}))= \frac{1}{|\det \partial \textbf{f}_i/\partial \textbf{x}_j| }\delta^{(n)}(\textbf{x}-\textbf{x}_0), \]
where the $\textbf{x}_0$'s are the zeros of $\textbf{f}(\textbf{x})$. Here $\textbf{f}=(H, D)^T$, $\textbf{x}=(R, p_R)^T$ and $\textbf{x}_0=(R^*, 0)^T$ with $\textbf{x}_0$ solving the constraints $H=0$ and $D=0$. That is,
\begin{equation} \delta(H) \delta(D)= \frac{1}{|\{H,D\}^*|}\delta(R-R^*) \delta(p_R-0).\end{equation}
Inserting this into (5.10), we obtain
\begin{eqnarray} && \int |\{D,H\}^*| \delta(D)\delta(H)\mathrm{d}R \mathrm{d}\psi \mathrm{d}\phi  \mathrm{d}p_R \mathrm{d}p_\psi  \mathrm{d}p_\phi \nonumber\\
&=&\int \delta(R-R^*)\delta(p_R) \mathrm{d}R \mathrm{d}\psi \mathrm{d}\phi  {d}p_R \mathrm{d}p_\psi \mathrm{d}p_\phi \nonumber\\
&=& \int \mathrm{d}\psi\mathrm{d}\phi\mathrm{d}p_\psi\mathrm{d}p_\phi.\end{eqnarray}
Hence, the Liouville measure $\mu$ on $T^*\mathcal{S}$ is given by
\begin{equation}\mu=\mathrm{d}\psi\mathrm{d}\phi\mathrm{d}p_\psi\mathrm{d}p_\phi.\end{equation}

It is a consequence of Faddeev's construction that $\mu$ is gauge-invariant. Moreover, since the reduced internal dynamics (the dynamics on $T^*S$) is Hamiltonian, it is stationary. That is, volume is conserved under the internal time evolution. 

To see this consider the Lie derivative $L_{X_\mathcal{H}} $ of the measure along the Hamiltonian vector field $X_\mathcal{H}$ on $T^*S$, the vector field connected to $\mathcal{H}$ from (4.15). The Lie derivative vanishes,
\begin{equation} L_{X_\mathcal{H}} \mu =0.\end{equation} 
Hence, $\mu$ is conserved under the time evolution connected to the Hamiltonian vector field $ {X_\mathcal{H}} $ on $T^*S$.
 
That the Lie derivative vanishes can be proven as follows. Note that $\mu = 1/2( \omega \wedge \omega)$, where $\omega=\mathrm{d}\psi\wedge \mathrm{d}p_\psi + \mathrm{d}\phi\wedge \mathrm{d}p_\phi$ is a symplectic two-form on $T^*S$. Since it is symplectic, it is closed, $\mathrm{d}\omega =0$. Using this and the fact that the vector field is Hamiltonian, i.e., $\omega (X_\mathcal{H}, \cdot)=\mathrm{d}\mathcal{H}$, we obtain: 
\begin{eqnarray}L_{X_\mathcal{H}}\mu&=& \frac{1}{2}(L_{X_\mathcal{H}}\omega \wedge \omega + \omega\wedge L_{X_\mathcal{H}}\omega )\nonumber\\
&=& \frac{1}{2}\big[(\mathrm{d}\omega)(X_\mathcal{H}, \cdot, \cdot) + \mathrm{d}(\omega(X_\mathcal{H}, \cdot))\big] \wedge \omega + \frac{1}{2}\omega \wedge\big[(\mathrm{d}\omega)(X_\mathcal{H}, \cdot, \cdot) + \mathrm{d}(\omega(X_\mathcal{H}, \cdot))\big]\nonumber\\
&=& \frac{1}{2}\big[0+ \mathrm{d}\circ \mathrm{d} \mathcal{H}\big] + \frac{1}{2}\big[\mathrm{d}\circ \mathrm{d} \mathcal{H} + 0\big] =0. \end{eqnarray}
Since the Lie derivative along the internal Hamiltonian vector field vanishes, the measure is conserved under the internal time evolution. Consequently, it is the appropriate measure for the statistical analysis on $T^*\mathcal{S}$.


\newpage

\begin{thebibliography}{11pt}

\bibitem{Arnold} Arnol'd, V. I. (1989): \textit{Mathematical Methods of Classical Mechanics.} Springer.

\bibitem{Barbour}Barbour, J., Koslowski, T., and Mercati, F. (2013): ``A gravitational origin of the arrows of time.'' ArXiv: 1310.5167 [gr-qc].

\bibitem{Barbour}Barbour, J., Koslowski, T., and Mercati, F. (2014): ``Identification of a gravitational arrow of time.'' Phys. Rev. Lett. 113:181101. ArXiv: 1409.0917 [gr-qc].

\bibitem{Barbour}Barbour, J., Koslowski, T., and Mercati, F. (2015): ``Entropy and the typicality of universes.''  ArXiv: 1507.06498 [gr-qc].

\bibitem{Faddeev} Faddeev, L. D. (1969): ``The Feynman Integral for Singular Lagrangians.'' \textit{Theoretical and Mathematical Physics} 1:1. 1-13

\bibitem{Iwai} Iwai, T. (1987): ``A geometric setting for classical molecular dynamics.'' \textit{Annales de l`I.H.P. Physique theorique} 47:2. 199-219.

\bibitem{Koslowski}Koslowski, T., Mercati, F., and Sloan, D. (2016): ``Through the Big Bang.'' ArXiv: 1607.02460 [gr-qc].

\bibitem{Marchal} Marchal, C. and Saari, D. (1976): ``On the final evolution of the n-body problem''. \textit{Journal of differential equations} 20. 150-186.

\bibitem{Marsden} Marsden, J., and Weinstein A. (1974): ``Reduction of symplectic manifolds with symmetries.'' \textit{Reports on Mathematical Physics} 5. 121-130.

\bibitem{Moeckel}Moeckel, R. (1981): ``Orbits of the three-body problem which pass infinitely close to triple collision.'' \textit{American Journal of Mathematics} 103: 6. 1323-1341.

\bibitem{Moeckel}Moeckel, R. (2007): ``Symbolic Dynamics in the planar three-body problem.'' \textit{Regular and Chaotic Dynamics} 12:5. 449-475.

\bibitem{Montgomery}Montgomery, R. (2002): ``Infinitely many syzygies.'' \textit{Arch. Rational Mech. Anal.} 164. 311-340.

\bibitem{Saari}Saari, D. G. (1984): ``Manifold structure for collisions and for hyperbolic-parabolic orbits in the n-body problem.'' \textit{Journal of differential equations} 55. 300-329.

\bibitem{Saari} Saari, D.: ``Central configurations - a problem for the twenty-first century.''  http://www.math.uci.edu/~dsaari/BAMA-pap.pdf


\bibitem{Sundman}Sundman, K. F. (1909): ``Nouvelles recherches sur le probleme des trois corps.'' \textit{Acta Soc. Sci. Fenn} 35. 1-27.



\end{thebibliography}
\end{document}